\documentclass[letterpaper]{JHEP3}

\pdfoutput=1

\newcommand{\tr}{\mathrm{Tr}}

\newcommand{\ba}{\begin{eqnarray}}
\newcommand{\ea}{\end{eqnarray}}

\usepackage{amsfonts}
\usepackage{amsmath}
\usepackage{amssymb}
\usepackage{cite}

\title{
Warm $p$-soup and near extremal black holes
}

\author{Takeshi Morita  \\
{\it
Department of Physics and Astronomy\\
 University of Kentucky, Lexington, KY 40506, USA
} \\
}

\author{Shotaro Shiba \\
{\it KEK Theory Center, High Energy Accelerator Research Organization (KEK) \\
1-1 Oho, Tsukuba, Ibaraki 305-0801, Japan } \\
}

\author{Toby Wiseman\\
{\it Theoretical Physics Group, Blackett Laboratory, Imperial College, London SW7 2AZ, UK } \\
}

\author{Benjamin Withers\\
{\it Mathematical Sciences and STAG Research Centre\\
University of Southampton, Highfield, Southampton SO17 1BJ, U.K.
 } \\
}

\preprint{\small KEK-TH-1689}

\date{November 2013}

\abstract{

We consider a model of $D$-dimensional supergravity coupled to elementary $p$-branes. 
We use gravitational arguments to deduce the low energy effective theory of $N$ nearly parallel branes. This is a $(p+1)$-dimensional scalar field theory, where the scalars represent the positions of the branes in their transverse space.
We propose that the same theory in a certain temperature regime describes a `soup' of strongly interacting branes, giving a microscopic description of near extremal black $p$-branes. 
We use natural approximations to estimate the energy density of this soup as a function of the physical parameters; $N$, temperature, brane tension and gravitational coupling.
We also characterise the horizon radius, measured in the metric natural to the branes, with the thermal vev of the scalars.
For both quantities we find agreement  with the corresponding supergravity black brane results.
Surprisingly, beyond the physical parameters, we are naturally able to reproduce certain irrational factors such as $\pi$'s. 
We comment on how these ideas may explain why black hole thermodynamics arises in gauge theories with holographic duals at finite temperature.

}

\begin{document}
\setlength{\baselineskip}{18pt}

%
\section{Introduction}
%

There has been considerable progress in the last twenty years in understanding black hole entropy from a microscopic perspective in string theory.
With sufficient charges certain supersymmetric string theory black holes may have entropy at zero temperature due to their highly degenerate vacuum structure. Counting BPS states at weak gravitational coupling has allowed matching with the Bekenstein-Hawking entropy starting in the work \cite{Strominger:1996sh}, and more recently allowing $\alpha'$ corrections to be reproduced \cite{Dabholkar:2004yr,Dabholkar:2005by}.
These methods have given spectacular precision matching, but are indirect in the sense that the entropy is computed in a weakly coupled regime, not describing black holes. This is argued to continue to strong gravitational coupling due to supersymmetry. 
However, being indirect, they do not elucidate a physical picture of the strongly coupled system.

Direct approaches have been made using the holographic description \cite{Maldacena:1997re,Itzhaki:1998dd} of near extremal black branes in terms of gauge theories. In particular these approaches have been used to understand situations where one is not counting entropy associated to ground state degeneracy, but rather the entropy arising strictly due to finite temperature and hence from non-supersymmetric dynamics. Analytic approaches for $0$-branes have been made using a method somewhat analogous to mean field \cite{Kabat:1999hp,Kabat:2000zv,Kabat:2001ve} although their accuracy is still unclear \cite{Lin:2013jra}.
Numerical finite temperature lattice approaches have been used and have shown consistency for $0$-branes \cite{Catterall:2007fp,Hanada:2007ti,Anagnostopoulos:2007fw,Catterall:2008yz} and some agreement also for $1$-branes \cite{Catterall:2010fx}. 
While these numerical  approaches provide powerful evidence that the gauge theory reproduces black hole behaviour at finite temperature in the appropriate temperature range where they describe dual black holes, and in principle can allow a wide variety of questions to be asked about these quantum black holes (see for example the very recent \cite{Hanada:2013rga}), there is no physical picture explaining \emph{why} black hole behaviour emerges.\footnote{
These theories have also been explored in a high temperature regime using numerical methods \cite{Asplund:2011qj,Riggins:2012qt,Asplund:2012tg,Berenstein:2013tya}, where the theory is not dual to black holes, but qualitatively similar behaviour has been observed. 
The issue of the IR instability was analytically studied in the high temperature regime \cite{Hotta:1998en,Mandal:2009vz}.
 Time dependent black hole physics has also recently been considered \cite{Iizuka:2013yla}.
}

The aim of this paper is to develop a simple physical model that underlies the black hole behaviour relevant for near extremal black holes in supergravity, string theory and holography. 
We will provide evidence that the near extremal thermodynamics of black branes in supergravity may be thought of as arising from a low temperature bound state of elementary branes. 
This is strongly gravitationally bound, and hence cannot be thought of as a weakly interacting gas. Rather we prefer the description as a warm soup of branes. The main results of this paper are to describe the classical theory governing these branes, and to argue that when it is put at finite temperature, the strongly interacting virialised behaviour appears to very naturally reproduce black hole behaviour including its thermodynamics and the relation between the black hole size and temperature.

We use the approach of \cite{Wiseman:2013cda,Morita:2013wla} (motivated by the work of \cite{Smilga:2008bt}) where the thermodynamics of near extremal $p$-branes was proposed to naturally arise from the dual gauge theory. However, this was not understood in terms of elementary branes, but from a gauge theory perspective. 
This work can be viewed as a refinement and rephrasing of the arguments in \cite{Wiseman:2013cda,Morita:2013wla} in purely gravitational terms. 
As we will discuss later, one important implication of our ideas is that the fact that holographic gauge theories exhibit black hole thermodynamic behaviour is a simple consequence of the fact that they reproduce the classical theory of gravitationally interacting branes.
This latter fact is of course a key piece of evidence for holography, and is very surprising. Whilst naively it is very surprising that they also reproduce black hole thermodynamics, our new view is that this is actually a simple consequence of gauge theories encoding brane dynamics and their interactions.

Although our study is motivated by superstring theory where the spacetime dimension $D = 10,11$, we will consider $p$-branes in general $D$ with $p \le D-4$. 
The fact that our arguments appear to apply for arbitrary $D$ may suggest that our microscopic description could apply independently of superstring theory, if a suitable $D$-dimensional supersymmetric quantum gravity exists.

\subsection{The difference between Schwarzschild and near extremal black holes}

Suppose we wished to have a microscopic description for the thermodynamics of a 4-d Schwarzschild black hole. The most naive thing we might try is to consider it to be a thermal bound state composed of $N$ gravitating particles, all with mass $m$, in flat space. Now consider the system in the Newtonian limit ie. at weak gravitational coupling and with the particles slowly moving. This gives rise to an action with pairwise interactions at leading order,
\begin{eqnarray}
\label{eq:sch}
S_{micro} = \int dt \, \frac{m}{2} \sum_{a=1}^N \partial_t{ \vec{x}_a } . \partial_t \vec{x}_a + \frac{\kappa^2 m^2}{4} \sum_{a<b} \frac{1}{4 \pi | \vec{x}_a - \vec{x}_b |}\,,
\end{eqnarray}
where $\kappa^2 = 16 \pi G_N$,  and $m$ is the mass of each particle, $\vec{x}_a$ is the spatial location of the $a$'th particle, with $a = 1, \ldots, N$.  Treating this system at finite temperature, we imagine that it forms a bound thermal gas, and then the virial theorem tells us,
\begin{eqnarray}
m \sum_{a=1}^N \langle \partial_t{ \vec{x}_a } . \partial_t \vec{x}_a \rangle \sim \kappa^2 m^2 \sum_{a<b} \langle \frac{1}{ | \vec{x}_a - \vec{x}_b |} \rangle \,.
\end{eqnarray}
Estimating the typical particle velocity as $v$, and separation as $x$, the virial theorem implies,
\begin{eqnarray}
\frac{ \kappa^2 M_{tot} }{x} \sim v^2\,,
\end{eqnarray}
where the total mass is $M_{tot} = N m$ in the Newtonian limit. Thus when the cloud is very diffuse (for example, a galaxy) with a scale $x \gg R_s$ with $R_s = \kappa^2 M_{tot}$ being the Schwarzschild radius, it behaves as a Newtonian gas. However, if we tried to make the cloud have a size smaller than $R_s$, we would simply learn that our Newtonian approximation ($v \ll 1$) had completely broken down. Hence we would not expect to usefully describe a Schwarzschild black hole in terms of an effective theory of massive slowly moving particles.

Now we consider a near extremal black hole in a 4 dimensional supergravity. 
A naive microscopic model for this black hole would be $N$ particles which have the mass $m$ as well as the same charge satisfying the BPS condition.
Then the above weakly interacting picture is slightly modified to the following, 
\begin{eqnarray}
\label{eq:4d}
S_{micro} = \int dt \,\frac{m}{2} \sum_{a=1}^N \partial_t{ \vec{x}_a } . \partial_t \vec{x}_a + \frac{\kappa^2 m^2}{4} \sum_{a<b} \frac{ | \partial_t{ \left( \vec{x}_a - \vec{x}_b\right) } |^4 }{4 \pi | \vec{x}_a - \vec{x}_b |} 
\end{eqnarray}
in the Newtonian limit. In this case making the same estimates, the virial theorem for a bound thermal gas tells us,
\begin{eqnarray}
\frac{ \kappa^2 M_{tot}}{x} \sim \frac{1}{v^2}\,.
\end{eqnarray}
The small velocity modification to the interaction totally changes the physics. Instead of the Newtonian limit ($v \ll 1$) applying to gas clouds with radii $x \gg R_s$, it now precisely appears to apply in the opposite limit, where $x \ll R_s$. Thus we might naively be led to the conclusion that we can treat a black hole simply as a Newtonian gas of particles. 
However we will see that these particles are strongly gravitationally coupled (in a sense that we will explain in section \ref{sec:estimates}) and thus while slowly moving, they are not in the Newtonian limit. Hence we regard the bound state as a soup rather than a gas.
We will find that natural estimates to extract the thermodynamics of this soup will lead us to reproduce the thermodynamic behaviour of near extremal black holes in the supergravity. Not only we will be able to reproduce parametric dependence on the particle mass $m$ (or tension for the brane cases), $\kappa^2$, $N$ and temperature, we will also be able to naturally reproduce certain irrational factors ($\pi$'s and sphere volumes).
In addition to estimating the black hole thermodynamics, we conjecture that the horizon location --- let us denote it $z_h$, although we will be careful in defining this later --- is related to the natural thermal vev of the positions of the elementary branes in the soup;
\begin{eqnarray}
z_h \sim x \sim \sqrt{ \frac{1}{N} \left\langle \vec{x}_a \cdot \vec{x}_a \right\rangle }\,.
\end{eqnarray}

Previously in the context of Matrix theory \cite{Horowitz:1997fr,Li:1997iz,Banks:1997tn,Li:1998ci} argued that certain M-theory black holes might be described by a thermal virialised bound state of elementary D0-branes following from the earlier work on the correspondence principle \cite{Horowitz:1996nw}.\footnote{
In fact in these works the black holes are not thought to be a bound states of individual D0-branes, but a bound state of ``zero-energy bound states of D0-branes'' \cite{Li:1998ci}. 
} 
While the black holes they discuss are not the same as the ones we discuss here\footnote{
In the case of $0$-branes we will discuss, our solutions can be lifted to M-theory.
There our solution is a homogeneous black string wrapping the M-theory circle, whereas \cite{Horowitz:1997fr,Li:1997iz,Banks:1997tn,Li:1998ci} study the localised black hole, which is a different branch of solutions.
We will discuss a phase transition between these two branches in the context of our effective moduli description in future work \cite{upcoming}. 
}
and the technicalities of those works are in detail quite different to our proposals here, a similar picture will emerge. \\

The paper is arranged as follows. Firstly in section \ref{sec:bh} we review the relevant aspects of supergravity $p$-brane black holes and their near extremal limit (more technical details are given in appendix \ref{app:sugra}). We then introduce the effective moduli theory that describes nearly parallel branes that interact gravitationally. In section \ref{sec:estimates} we argue that placing this effective theory at finite temperature naturally reproduces the near extremal thermodynamics of the black $p$-branes and the behaviour of their horizon size. Finally in section \ref{sec:discussion} we conclude with a discussion. In particular we emphasise that if our ideas are correct, they provide a simple explanation for why black hole thermodynamic behaviour is seen in holographic gauge theories.

%
\section{Supergravity coupled to elementary branes \label{sec:bh}}
%

We now consider a theory of dilaton gravity that admits extremal charged $p$-brane solutions, and its coupling to elementary $p$-branes. 
We require that the extremal branes satisfy a `no-force' condition, and hence that multi-centred asymptotically flat extremal solutions may be found. Whilst we will not consider fermionic content of this theory, such behaviour naturally arises in the supergravity context.

We take the total spacetime dimension to be $D$, and consider $p$-branes with $n$ transverse dimensions, so that $D = 1 + p + n$.
The theory will encompass all the string and M-theory branes (that preserve 16 supercharges) and where only one brane variety is present, though we will not limit our attention to such cases. In particular it includes the IIA/B string theory D$p$-branes ($D = 10$ and $n = 9-p$), and the M-theory M$2$-brane ($D=11$, $p=2$ and $n=8$) and M$5$-brane ($D=11$, $p=5$ and $n=5$). The remaining cases of the F$1$ and NS$5$ branes are related by electric-magnetic duality to the D$1$ and D$5$. For $D < 10$ the theory can be obtained by a consistent reduction of the $D = 10, 11$ cases\cite{Duff:1994an}.

We start with metric $g_{MN}$, dilaton $\phi$ and $1+p$-form potential $A_{1+p}$, so $F_{2+p} = d A_{1+p}$, governed by the action;
\begin{eqnarray}
\label{eq:sugra}
I = \frac{1}{2 \kappa^2} \int d^D x \sqrt{-g} \left( R - \frac{1}{2} (\partial \phi)^2 - \frac{1}{2 (p+2)!} e^{a\, \phi} F^2_{2+p} \right) 
\end{eqnarray}
where the constant $a$ must take the particular value
\begin{eqnarray}
a^2 = 4 - \frac{2 (p+1)(n-2) }{D - 2} 
\end{eqnarray}
in order to have asymptotically flat solutions, and so that extremal solutions obey a `no-force' condition when static. 
The gravitational coupling is $\kappa^2$. An extremal solution saturates the BPS bound $Q = \sqrt{2} \kappa M$, where $Q$ and $M$ are charge and mass density respectively. 
We take the elementary brane in this theory to have tension $\mu$. The action for an elementary brane with world volume coordinates $\xi^\nu$, $\nu = 0, \ldots, p$ and embedding described by the functions $x^M = Z^M( \xi^\nu )$ is,
\begin{eqnarray}
\label{eq:elementary}
S &=& - \mu \left( \int d^{1+p} \xi \sqrt{ - \gamma}  + \int A \right) 
\end{eqnarray}
where $A$ is the pull back of $A_{1+p}$ to the world volume of the branes, and,
\begin{eqnarray}
\gamma_{\mu\nu} = \partial_\mu Z^M \partial_\nu Z^N \left( g_{MN} e^{\frac{a}{p+1} \phi} \right)
\end{eqnarray}
so $\gamma_{\mu\nu}$ is the metric on the world volume induced from the spacetime metric $g_{MN} e^{\frac{a}{p+1} \phi}$. 
An elementary brane saturates the BPS bound with charge density $q = \sqrt{2} \kappa \mu$. Hence the number of elementary branes $N$ (which is quantised) for a general solution will be,
\begin{eqnarray}
N = \frac{Q}{q} =  \frac{Q}{ \sqrt{2} \kappa \mu}.
\end{eqnarray}
For an extensive review of this theory and its solutions see \cite{Duff:1994an}.

We note that we have referred to our actions \eqref{eq:sugra} and
\eqref{eq:elementary} as those of supergravity. However,  we
have not presented the
 fermions (and world volume gauge fields if they are required by supersymmetry),
and clearly for arbitrary $D$ and $p$ we will not be able to find a
supersymmetric embedding of this bosonic theory.
In fact we will see that the bosonic theory is enough to explore the
black brane thermodynamics.
It is presumably important that our theory does arise in a
supersymmetric context, although this may be more important for a
fundamental quantum description of the gravity theory,  than for the
arguments we make here.

\subsection{ Effective theory for gravitationally interacting elementary branes}

We will derive an effective action for $N$ interacting $p$-branes 
in the above model, which for $D=4$ and $p=0$ will look like the one in the introduction (\ref{eq:4d}).
To evaluate the gravitational forces between the branes, we start considering extremal $p$-brane solutions.

Due to the no-force condition, vacuum solutions consist of any number of parallel branes, displaced in the transverse space. The solution for $N$ elementary brane sources with tension
and couplings as above is then (in Einstein frame metric),
\begin{eqnarray}
\label{eq:multimetric}
ds^2 &=& H^{-\frac{n-2}{D-2}} \left( - dt^2 +  dx^i dx^i  \right) +  H^{\frac{{p+1}}{D-2} } \left( \delta_{mn} dZ^m dZ^n \right) \,, 
\end{eqnarray}
where,
\begin{eqnarray}
\label{eq:multiharmonic}
H = 1 + \sum_{a=1}^{N} \frac{2 \kappa^2 \mu_a}{(n-2)\, \Omega_{n-1}}  \frac{1}{ | Z^m - z^m_{(a)} |^{n-2}}
\end{eqnarray}
for $N$ parallel branes with positions $z_{(a)}^m$ in the transverse space, with $a = 1, \ldots, N$, each with tension $\mu_a$. Here $\Omega_{n-1}$ is the volume of a unit $(n-1)$-sphere,
\begin{eqnarray}
 \Omega_{n-1} \equiv \frac{ 2 \pi^{\frac{n}{2}} }{ \Gamma\left( \frac{n}{2} \right) }
\end{eqnarray}
and $H$ is a harmonic function in the transverse space with delta-function sources at the positions $z_{(a)}^m$.

Consider a single extremal brane at the origin in the transverse space, with tension $\mu_1$. In this case, \eqref{eq:multiharmonic} reduces to, 
\begin{eqnarray}
H = 1 +  \frac{2 \kappa^2 \mu_1}{(n-2)\, \Omega_{n-1}}  \frac{1}{ | Z^m |^{n-2}}\,.
\end{eqnarray}
Now consider a second brane with tension $\mu_2$ that is at position $Z^m(t,x)$, and we use the usual $\mathbb{R}^n$ vector notation for the transverse space so $Z^m = \vec{Z}$. Consider this brane as a probe in the background of the first. From \eqref{eq:elementary} the action for this probe brane in the extremal background is, 
\begin{eqnarray}
S_{probe} =  - \mu_2 \int dt dx^p \frac{1 }{H} \left( \sqrt{ - \det\left( \eta_{\mu\nu} + H  \partial_\mu \vec{Z} \cdot \partial_\nu \vec{Z}   \right)  }  - 1 \right)\,,
\end{eqnarray}
which may be expanded in powers of $\kappa$, as,
\begin{eqnarray}
S_{probe} &=&   - \int dt dx^p \mu_2 \left( \sqrt{ - \det \mathbf{A}  }  - 1 \right)  \\
&& \qquad  +\,  \kappa^2 \frac{2 \mu_1 \mu_2}{(n-2)\, \Omega_{n-1}} \frac{1}{ | \vec{Z} |^{n-2} }   \left(  1 + \sqrt{ - \det \mathbf{A}  } \left(  \frac{1}{2} \tr\left( \mathbf{A}^{-1} \mathbf{B} \right) - 1 \right) \right) + O(\kappa^4)\,, \nonumber 
\end{eqnarray}
where $A_{\mu\nu} = \eta_{\mu\nu} + \partial_\mu \vec{Z} \cdot \partial_\nu \vec{Z}$ and $B_{\mu\nu} = \partial_\mu \vec{Z} \cdot \partial_\nu \vec{Z}$.
The first term is due to the tension of the brane and the latter term is due to inter-brane interaction by  classical supergravity dilaton-graviton-form field exchange. We may rewrite its form as, 
\begin{eqnarray}
S_{probe} &=&  - \int dt dx^p \mu_2 \left( \sqrt{ - \det \mathbf{A}  }  - 1 \right)  \\
&& \qquad  +\,   \frac{2 \kappa^2 \mu_1 \mu_2}{(n-2)} h( | \vec{Z} | )  \left(  1 + \sqrt{ - \det \mathbf{A}  } \left(  \frac{1}{2} \tr\left( \mathbf{A}^{-1} \mathbf{B} \right) - 1 \right) \right) + O(\kappa^4)\,, \nonumber 
\end{eqnarray}
where, writing the transverse space $\mathbb{R}^n$ as, $ds^2 = dr^2 + r^2 d \Omega_{n-1}^2$,
then $h(r)$ is the harmonic function which results from a unit strength delta function source,
\begin{eqnarray}
h( r ) = \frac{1}{ \Omega_{n-1} } \frac{1}{r^{n-2}}\,.
\end{eqnarray}
We will be interested in the case where the branes are weakly curved in the sense that $\left| \partial_\mu \vec{Z} \right| \ll 1$, so that we may approximate, 
\begin{eqnarray}
 \sqrt{ - \det\left( \eta_{\mu\nu} + \partial_\mu \vec{Z} \cdot \partial_\nu \vec{Z}   \right)  }  - 1  \simeq  \frac{1}{2} \eta^{\mu\nu} \partial_\mu \vec{Z} \cdot \partial_\nu \vec{Z} + \ldots\,.
\end{eqnarray}
This approximation is analogous to the slow moving requirement of the Newtonian approximation for the $p=0$ particle model (\ref{eq:4d}).
In such a case the branes are nearly parallel and we may interpret $\vec{Z}(t, x)$ as a physical modulus describing the separation between the two branes in their mutual transverse space, the source brane being at the origin, and the probe at $\vec{Z}$.

Now consider $N$ elementary branes, each with tension $\mu$. Let us take these branes to be parallel, and then consider long wavelength fluctuations. 
While normally the interaction between two branes in general is non-local viewed from their world volume, and mediated by the supergravity $D$-dimensional propagator, in such a long wavelength limit we expect an effective action for these $N$ branes which is \emph{local}, and involves the propagator in the transverse dimensions to the branes.
We think of the branes as sharing common world volume coordinates, and describe their positions in the transverse space by the scalars $\vec{\Phi}_a$ with $a = 1, 2, \ldots, N$. One may think of these as `moduli' or Nambu-Goldstone modes due to the branes breaking the translation invariance in the transverse space.
From the above probe calculation we can simply deduce the $(p+1)$-dimensional
theory governing these $N$ moduli fields to first sub-leading order in $\kappa^2$ as,
\begin{eqnarray}
 S_{moduli} &=&   -\int dt dx^p \frac{\mu}{2} \sum_{a}   \partial_\mu \vec{\Phi}_a \cdot \partial^\mu \vec{\Phi}_a  \left( 1 + O\left( ( \partial \vec{\Phi} )^2 \right) \right)  \nonumber \\
 && 
   +\,  \frac{\mu^2 \kappa^2}{4 (n-2) \Omega_{n-1}} \int dt dx^p \sum_{a<b}^{}  \frac{ 2 \left( \partial_\mu \vec{\Phi}_{ab} \cdot \partial_\nu \vec{\Phi}_{ab} \right)^2 -  \left( \partial_\mu \vec{\Phi}_{ab} \cdot \partial^\mu \vec{\Phi}_{ab} \right)^2 }{ \left| \vec{\Phi}_{ab} \right|^{n-2} }  \left( 1 + O\left( ( \partial \vec{\Phi} )^2 \right) \right)   \nonumber \\
 && 
  +\, O( \kappa^4  ) \,,
\label{eq:eff-action}
 \end{eqnarray}
  where we have defined the separation between the $a$'th and $b$'th brane in the transverse space as,
 \begin{eqnarray}
 \vec{\Phi}_{ab} \equiv \vec{\Phi}_a - \vec{\Phi}_b\,.
 \end{eqnarray}
We term this the `moduli theory' governing the low energy dynamics of the $N$ branes away from their parallel vacuum configurations. 
Note that in the $D=4$ and $p=0$ case, this model is identical to the model (\ref{eq:4d}) for the 4 dimensional supergravity once we identify $m=\mu$.
 
The first term in the $\kappa$ expansion is due to the fluctuations of the $N$ branes, and we have assumed they are weakly curved so that $( \partial \vec{\Phi} )^2 \ll 1$. Hence at leading order in the $\kappa$ expansion we ignore terms involving four derivatives or higher.
The second term is due to the pairwise dilaton-graviton-form mediated interaction and again, since we take $( \partial \vec{\Phi} )^2 \ll 1$ higher derivative corrections to this are neglected. At this order in $\kappa$ such terms begin at six derivatives.
At higher orders in $\kappa$ we will expect more complicated gravitational interactions between the branes. 
We expect that at order $\kappa^{2l}$ there will be terms due to exchange of $l$ gravitons between $l+1$ of the $N$ branes of the schematic form,
\begin{eqnarray}
\label{eq:onshell}
 S_{on-shell, 2l} &=&  c_l \mu  \left(  \mu \kappa^2 \right)^{l} \int dt dx^p \sum_{a_0, a_1, \ldots , a_l}^{}   \left( \partial \vec{\Phi} \right)^{2 (l+1)}  h\left( \left| \vec{\Phi} \right| \right)^{l} \nonumber \\
  & = & c_l \mu \left( \frac{ \mu \kappa^2}{\Omega_{n-1}} \right)^{l} \int dt dx^p \sum_{a_0, a_1, \ldots , a_l}^{}  \frac{ \left( \partial \vec{\Phi} \right)^{2 (l+1)} }{ \left| \vec{\Phi} \right|^{l ( n - 2 )} }\,,
 \end{eqnarray}
where $c_l$ are $l$-dependent constants not containing any of $N,\mu,\kappa$, factors of $\pi$ or sphere volumes $\Omega_{n-1}$. In the classical supergravity coupled to branes such contributions are `on-shell'.
However, there will also be more complicated terms involving classical exchange of more than $l$ gravitons which themselves interact in the bulk. Such terms in classical effective theory are `off-shell' contributions. It would be interesting to understand the structure of these terms further. However, we leave that for future work.

In summary, we have the description of the dynamics of nearly parallel weakly curved elementary branes in flat space, including the leading order interaction mediated by graviton exchange. Later in the paper we will argue that putting this theory at finite temperature, and requiring the gravitational interaction to be strong so that the branes are bound together, will naturally reproduce the near extreme thermodynamics and behaviour of black $p$-brane solutions in our supergravity theory. We now review these black holes and their behaviour.

\subsection{ Near extremal black hole solutions}

As we detail in the Appendix \ref{app:sugra} we may find asymptotically flat solutions representing $N$ parallel $p$-branes. For these we may take the decoupling limit. Starting with the near extremal solutions, one focuses on the geometry of the black hole throat, and as described in the appendix, find this geometry is described by the metric,
\begin{eqnarray}
\label{eq:metric-bb}
ds^2 &=& \left( \frac{(n-2)\, \Omega_{n-1}}{2 \kappa^2 \mu N}   z^{n-2} \right)^{\frac{n-2}{D-2}} \left( - f dt^2 +  dx^i dx^i  \right) +  \left( \frac{(n-2)\, \Omega_{n-1}}{2 \kappa^2 \mu N}   z^{n-2} \right)^{- \frac{{p+1}}{D-2} } \left( \frac{1}{f} dz^2 + z^2 d\Omega^2_{n-1} \right) \nonumber \\
f & = & 1 - \left( \frac{z_h}{z} \right)^{n-2} 
\end{eqnarray}
and in addition the dilation and $(1+p)$-form have non-trivial $z$-profiles too. Here $d \Omega^2_{n-1}$ is the line element on the round unit $(n-1)$-sphere.

Notice that in the case of D$3$ branes ($p=3$, $n = 6$), M$2$ brane ($p=2$, $n=8$) or M$5$ branes ($p=5$, $n=5$) then $a = 0$ and hence the dilaton decouples from the gauge field, and it is constant in the solutions above. Correspondingly the geometries factorise in these cases to the familiar products $AdS_5 \times S^5$, $AdS_4 \times S^7$ and $AdS_7 \times S^4$ respectively.

The radial coordinate $z$ has conventional units of length. In this radial coordinate the position of the horizon is located at $z = z_h$. We term $z_h$ the `horizon location'. The quantity $z_h$ may appear simply to be a property of the particular coordinate system adopted above. However, the form of the elementary brane coupling to the gravity fields in fact gives it physical significance as we shall now discuss.
Firstly observe that the metric above factors into the warped product of the brane world volume,
\begin{eqnarray}
ds^2_{brane} = g^{brane}_{\mu\nu} dx^\mu dx^\nu = - f dt^2 + dx^i dx^i
\end{eqnarray}
 with the $n$-space transverse to the branes with metric,
\begin{eqnarray}
ds^2_{transverse} = g^{transverse}_{mn} dZ^m dZ^n = \frac{1}{f} dz^2 + z^2 d \Omega^2_{n-1}\,,
\end{eqnarray}
 which is asymptotic to the flat Euclidean space $\mathbb{R}^{n}$. This transverse metric has an important physical significance. Consider an elementary $p$-brane moving in the extremal case of the background above so $z_h = 0$, where now, $g^{brane}_{\mu\nu} =\eta_{\mu\nu}$, and,
\begin{eqnarray}
ds^2_{transverse} = dz^2 + z^2 d \Omega_{n-1}^2 = \delta_{mn} dZ^m dZ^n
\end{eqnarray}
is the Euclidean metric.
Take the brane to have position $Z^m(t,x)$, and we use the usual $\mathbb{R}^n$ vector notation for the transverse space so $Z^m = \vec{Z}$. From \eqref{eq:elementary} the action for this probe brane in the extremal background is,
\begin{eqnarray}
S_{probe} =  - \mu \int dt dx^p \frac{ | \vec{Z} |^{n-2}  }{C} \left( \sqrt{ - \det\left( \eta_{\mu\nu} + C \frac{ \partial_\mu \vec{Z} \cdot \partial_\nu \vec{Z}   }{ \left|  \vec{Z} \right|^{n-2} } \right)  }  - 1 \right)
\end{eqnarray}
with $C ={2 N \kappa^2 \mu }/\left( ( n - 2) \Omega_{n-1} \right)$ 
which expanded to second order in derivatives is,
\begin{eqnarray}
\label{eq:probe-z}
S_{probe} \simeq  - \frac{\mu}{2} \int dt dx^p  \eta^{\mu\nu} \partial_\mu \vec{Z} \cdot  \partial_\nu \vec{Z} + \ldots\,.
\end{eqnarray}
More generally, in the non-extremal case, one finds the two derivative term (note that now there is also a zero derivative potential term too),
\begin{eqnarray}
\label{eq:probegeneral}
S_{probe, 2 \, deriv} \simeq  - \frac{\mu}{2} \int dt dx^p  g_{brane}^{\mu\nu} \partial_\mu Z^m   \partial_\nu Z^n g^{transverse}_{mn}\,.
\end{eqnarray}
Hence the transverse metric $g^{transverse}_{mn}$ is physically defined as the metric that governs probe brane motion. More formally it is the superspace metric for a probe brane action.
We may then conclude that the quantity $z_h$ is indeed physical, and gives the (areal) radius of the horizon as measured by a brane probe. Roughly speaking, it is the size of the horizon as seen by the branes (rather than by the $D$-dimensional Einstein metric).

The mass density, $M$, and temperature $T$ of the decoupling limit of the black $p$-brane solution is calculated from the asymptotically flat solutions from the usual ADM mass and horizon area before taking the decoupling limit, and then considering these quantities in the near extreme or decoupling limit. 
One finds the energy density above extremality in the decoupling limit, $\epsilon_{decoupling}$ is given in terms of the horizon position $z_h$ as,
\begin{eqnarray}
\label{eq:erg}
\epsilon_{decoupling} & = & \frac{n}{4} \frac{\Omega_{n-1}}{\kappa^2}  z_h^{n-2}\,,
\end{eqnarray}
and this horizon location depends on temperature as,
\begin{eqnarray}
\label{eq:zh}
z_h = \left( \frac{ 2^5  \pi^2 N T^2 \mu \kappa^2 }{(n-2)^3 \Omega_{n-1} } \right)^{\frac{1}{n-4}}\,.
\end{eqnarray}
From the two equations above we determine energy density in terms of temperature and may then compute the entropy density,
\begin{eqnarray}
\label{eq:sugraentropy}
T \, s_{decoupling} & = & 2^{\frac{4 n - 6}{n - 4}} \left( n-2 \right)^{- \frac{2 (n - 1)}{n - 4} } \left(  N \, \pi^2 \, T^2   \right)^{\frac{n-2}{n-4}}  \left( \frac{ \kappa^2  \mu^\frac{n-2}{2}   }{\Omega_{n-1}} \right)^{\frac{2}{n - 4}}\,.
\end{eqnarray}
The entropy density and horizon location give two measures of the `size' of the horizon. The former, via the Bekenstein formula, relates to the spatial horizon volume as measured by the $D$-dimensional metric. The latter is the horizon radius as measured by the branes themselves. We see that these two different measures have rather different dependences on the physical parameters in the theory. A key result of this work is that we will naturally be able to estimate both these dependences from the proposed microscopic effective theory.

As we show in the appendix, corrections to the near extremal limit are controlled by the dimensionless quantity, 
\begin{eqnarray}
\label{eq:omega}
\omega = \left(\Omega_{n-1}^{-1} N \kappa^2 \mu \pi^{n-2}T^{n-2} \right)^{\frac{2}{n-4}}\,,
\end{eqnarray}
so that, for example, the corrections to energy density are,
\begin{eqnarray}
\label{eq:decoupling}
\epsilon  & = & \epsilon_{decoupling}  \left( 1 + k \,  \omega +  O\left( \omega^2 \right) \right) \,, 
\end{eqnarray}
for a constant $k$,
\begin{eqnarray}
k = \frac{ 2^{\frac{2 (1+n)}{n-4}} (n-2)^{\frac{2 (1-n)}{n-4}} (3 n - 8)  }{n-4}.
\end{eqnarray}
The near extreme, or decoupling limit is given by the condition $\omega \ll 1$.
Note that the fidelity of this condition is such that it correctly accounts for factors of $\pi$ and the sphere volume which arise in deviations from extremality \eqref{eq:decoupling} -- this will be important later for our estimates which track such factors.

%
\section{Estimates for the thermal moduli theory \label{sec:estimates}}
%

We now consider the moduli theory (\ref{eq:eff-action}) at finite temperature.
We will use the Euclidean time formalism, so $\tau = i t$ and $\tau$ is periodic with period $\beta = 1/T$.
We assume that the branes compose a thermal bound state and estimate the thermodynamics through the virial theorem.
Then this estimation with natural approximations will lead us to conclude that the bound state is strongly coupled and the thermodynamics is consistent with that of near extremal black branes in  supergravity.
The estimation method we employ here is based on that used in \cite{Wiseman:2013cda,Morita:2013wla} in the context of the moduli theory of maximally supersymmetric gauge theories and recovering dual black hole physics, following earlier work \cite{Smilga:2008bt}. We will discuss this relation to the moduli theory of gauge theories in the later discussion in section \ref{sec:discussion}. However our perspective here is purely gravitational, based on the classical moduli theory of our nearly parallel, gravitationally interacting branes, and putting it at finite temperature.

By applying the virial theorem\footnote{
Here we are employing the virial theorem not only in the quantum mechanical cases ($p=0$) but also for extended worldvolumes ($p>0$).  
The virial theorem in this field theory context is discussed in Landau-Lifshitz \cite{Landau-vol3}.
} to the moduli theory (\ref{eq:eff-action}), we have up to rational numerical constants, that the energy density $\epsilon$ is related to the vevs of the kinetic and interaction terms as,
\begin{eqnarray}\label{eq:bigvirial}
\epsilon \sim \frac{\mu}{2}  \sum_{a} \Bigl\langle  \left( \partial_\mu \vec{\Phi}_a \right)^2 \Bigr\rangle \sim  \frac{\mu^2 \kappa^2}{\Omega_{n-1}} \sum_{a<b}^{}  \Bigl\langle  \frac{ \left( \partial_\mu \left( \vec{\Phi}_a -  \vec{\Phi}_b \right) \right)^4 }{ \left| \vec{\Phi}_a -  \vec{\Phi}_b \right|^{n-2} }  \Bigr\rangle\,.
\end{eqnarray}
We can use natural parametric estimates to understand qualitative features of the thermal behaviour.
We emphasise that in the following we use $\sim$ to denote equality including dependence on physical parameters $N, T, \mu, \kappa$, but also including all factors of $\pi$'s and sphere volumes $\Omega_{n-1}$.
 Firstly we assume that there is a dominate scale, $\Phi$, controlling the behaviour of the moduli fields so that we may estimate;
\begin{eqnarray}
\vec{\Phi}_a \sim \vec{\Phi}_a - \vec{\Phi}_b \sim \Phi \,.
\end{eqnarray}
Then all the scalars, and their differences, are of the same magnitude denoted by $\Phi$. 
Likewise we estimate the derivatives of the scalars, and their differences, also to have the same parametric magnitudes,
\begin{eqnarray}
\partial_\mu \Phi_a \sim \partial_\mu \left( \Phi_a - \Phi_b \right) \sim \partial \Phi \,,
\end{eqnarray}
where $\partial \Phi$ gives this scale. Thirdly, being at large $N$ we estimate,
\begin{eqnarray}
\sum_a \sim N \; , \quad \sum_{a<b} \sim N^2\,.
\end{eqnarray}

%
\subsection{Horizon size} 
%

The scalars $\vec{\Phi}_a$ describe how the branes move in the transverse space, and hence we expect the thermal vev of the scalars to give the size of the brane configuration in this transverse space, as measured in the flat metric on $\mathbb{R}^n$.

We now make a further key assertion. Based on the fact that $z_h$ is the radius of the black hole in the transverse space \emph{as measured by the branes themselves}, and that the $\vec{\Phi}_a$ are the coordinate positions of the branes in \emph{canonical coordinates} in the transverse $\mathbb{R}^n$,  we claim that the thermal vev of the scalars is given by,
\begin{eqnarray}
z_h^2 \simeq \frac{1}{N} \left\langle \left| \vec{\Phi}_a \cdot \vec{\Phi}_a \right| \right\rangle \,,
\end{eqnarray}
where the equality is accurate up to constant factors which do not involve any of $N,\mu,\kappa,\pi$ or $\Omega_{n-1}$. 
Then using our estimates above we find simply,
\begin{eqnarray}
\label{eq:zhest}
z_h \sim \Phi\,.
\end{eqnarray}

%
\subsection{Energy} 
%
Using our estimates we approximate the left-hand side of the virial theorem \eqref{eq:bigvirial} as, 
\begin{eqnarray}
\frac{\mu}{2}  \sum_{a} \Bigl\langle  \left( \partial_\mu \vec{\Phi}_a \right)^2 \Bigr\rangle \sim \mu N \left( \partial \Phi \right)^2\,,
\end{eqnarray}
and the right-hand side as,
\begin{eqnarray}
\frac{\mu^2 \kappa^2}{\Omega_{n-1}} \sum_{a<b}^{}  \Bigl\langle  \frac{ \left( \partial_\mu \left( \vec{\Phi}_a -  \vec{\Phi}_b \right) \right)^4 }{ \left| \vec{\Phi}_a -  \vec{\Phi}_b \right|^{n-2} }  \Bigr\rangle \sim\frac{\mu^2 \kappa^2}{\Omega_{n-1}}  N^2 \frac{ \left( \partial \Phi \right)^4 }{  \Phi^{n-2} }\,,
\end{eqnarray}
and equating these we may deduce the scale $(\partial \Phi)$ as,
\begin{eqnarray}
\label{eq:derivest}
 (\partial \Phi)^2 \sim \frac{1}{N \mu \kappa^2} \Omega_{n-1} \Phi^{n-2}\,.
\end{eqnarray}
Hence given that $\epsilon \sim \mu  \sum_{a} \Bigl\langle  \left( \partial_\mu \vec{\Phi}_a \right)^2 \Bigr\rangle$ we obtain,
\begin{eqnarray}
\label{eq:energy}
\epsilon \sim  \frac{1}{\kappa^2} \Omega_{n-1} \Phi^{n-2}\,.
\end{eqnarray}
Using the horizon size estimate \eqref{eq:zhest} so $\Phi \sim z_h$, we see that we reproduce the supergravity energy density relation in \eqref{eq:erg} up to a rational number.
\\

%
\subsection{Demonstration of strong coupling\label{higherOrderTerms}} 
%

Here we  estimate the on-shell higher order $\kappa$ terms as in \eqref{eq:onshell}.
  Using,
 \begin{eqnarray}
 \sum_{a_0, a_1 , \ldots , a_l} \sim N^{l+1}\,,
\end{eqnarray}
then we may estimate,
\begin{eqnarray}
 S_{on-shell, 2l}  & = & c_l \mu  \int d\tau dx^p \left( \partial \vec{\Phi} \right)^{2} \sum_{a_0, a_1, \ldots , a_l}^{} \left( \frac{ \mu \kappa^2}{\Omega_{n-1}} \frac{ \left( \partial \vec{\Phi} \right)^{2 } }{ \left| \vec{\Phi} \right|^{ ( n - 2 )} }  \right)^{l}
 \nonumber \\ & \sim & 
 \beta   \int dx^p N \mu \left( \partial \vec{\Phi} \right)^{2} \left( \frac{ \mu \kappa^2}{\Omega_{n-1}} \frac{ N \left( \partial \vec{\Phi} \right)^{2} }{ {\Phi}^{ ( n - 2 )} }  \right)^{l}\, .
 \end{eqnarray}
 Now using our estimate for $\Phi$ (\ref{eq:derivest}) we have that the term in the large brackets that controls the relative size of terms for different $l$,
 \begin{eqnarray}
 \frac{ \mu \kappa^2}{\Omega_{n-1}} \frac{ N \left( \partial \vec{\Phi} \right)^{2} }{ {\Phi}^{ ( n - 2 )} }  \sim 1
 \end{eqnarray}
 and hence the terms $ S_{on-shell, 2l} $ are all parametrically equal for any $l$. 
 This means that all higher interactions contribute equally and the bound state is strongly coupled (a soup of $p$-branes).

We started from the assumption that the thermal theory governed a bound state, and hence obeyed a virial theorem by considering only $l=0$ and $l=1$. However, we see this is actually consistent with all the on-shell interactions between the branes becoming strongly coupled for any $l$.

%
\subsection{Temperature} 
%

We now consider the temperature dependence of the bound state.
If the system is a weakly coupled gas then it might make sense to discuss a `virial temperature', i.e. $E_{tot} \sim N T_{\text{virial}} $ for $N$ particles.
However, as explained in section \ref{higherOrderTerms}, the bound state is strongly coupled.
We now make our key physical assumption for this strongly coupled regime, namely that it is the temperature that controls the scale of derivatives in the theory, so that,
\begin{eqnarray}
\label{eq:del-T}
\partial \Phi \sim \pi T \Phi \,.
\end{eqnarray}
 Whilst there is another dimensionful parameter\footnote{
  In the canonical normalisation,
 $ \kappa^2 \mu^{(n-2)/2} $ is the unique dimensionful parameter of the moduli theory (\ref{eq:eff-action}).} $ \kappa^2 \mu^{(n-2)/2} $, we have assumed that the derivatives scale with the 
 thermal scale $T$. 
 This is a natural scaling for a massless free field, or a theory at low energies where there is no mass gap and hence some IR scaling behaviour. 
 We will later see that this assumption reproduces the black brane dynamics consistently. Indeed in the appendix \ref{app:scaling} we may straightforwardly see a classical scaling behaviour inherent in our moduli action, and we may view our derivative estimate as being equivalent to considering the thermal behaviour to be in this scaling regime. 
  
A possible surprise at this point is the presence of the `$\pi$' above. We insert this as given a mode expansion on the Euclidean thermal circle of some field $\psi$,
\begin{eqnarray}
\label{eq:pi}
\psi(\tau, x) \sim \sum_n e^{2 \pi T n i \tau} \psi_n(x)\,,
\end{eqnarray}
it is clear the natural quantity is $\pi T$.\footnote{
We also note that if we evaluate $\langle \phi^2 \rangle$ and $\langle (\partial_\mu  \phi)^2 \rangle$ in a four dimensional free massless scalar theory, we can explicitly show $\langle (\partial_\mu  \phi)^2 \rangle/\langle \phi^2 \rangle \propto (\pi T)^2$, agreeing with the estimate $\partial_\mu \sim \pi T$. }
Naively we would never expect the parametric estimation we are going to perform to account for factors of $\pi$ --- nonetheless we shall see that inserting this natural factor of $\pi$ here does indeed correctly account for all the transcendental dependence of the thermal behaviour of the theory. 

From \eqref{eq:derivest} we now derive,
\begin{eqnarray}
\label{eq:derivest2}
 ( \pi T \Phi)^2 \sim \frac{1}{N \mu \kappa^2} \Omega_{n-1} \Phi^{n-2}\,,
\end{eqnarray}
and hence conclude,
\begin{eqnarray}
\label{eq:phi}
\Phi \sim \left( \frac{ \pi^2 N T^2 \mu \kappa^2 }{ \Omega_{n-1} } \right)^{\frac{1}{n-4}}
\end{eqnarray}
for $n \neq 4$.
Again we see, using \eqref{eq:zhest}, that this is in precise agreement with the corresponding  supergravity relation \eqref{eq:zh}.

Here the case $n=4$ is exceptional and it exhibits Hagedorn behaviour.
In equation (\ref{eq:derivest2}), $\Phi$ cancels for $n=4$ and, rather than the scalars being determined parametrically in terms of the temperature, instead the temperature is fixed at the Hagedorn temperature $T_H$,
\begin{eqnarray}
T_H & \sim & \left( \frac{ \Omega_{3} }{ N  \mu \kappa^2 \pi^2} \right)^{ \frac{1}{2} }, \qquad n = 4\,.
\end{eqnarray}
The scale $\Phi$ is not determined and remains as a free parameter of the system.
The Hagedorn behaviour is consistent with that of the black brane solutions in supergravity as presented in Appendix \ref{app:hagedorn} with the Hagedorn temperature (\ref{eq:TH})  consistently reproduced.
In  superstring theory, Hagedorn behaviour is observed in the $n=4$ cases of black D5 and NS5-branes \cite{Callan:1992rs, Maldacena:1997cg}, in the decoupling limit.
\\

%
\subsection{Entropy} 
%

Having obtained estimates for the energy and the temperature, using the first law an estimate of the entropy density of the black hole gives $s\sim \epsilon/T$. The estimates for these former quantities agree with the corresponding supergravity black hole values. Needless to say the entropy density is also in agreement with \eqref{eq:sugraentropy}, but we take this opportunity to reiterate that our estimates are microscopic in origin. 

%
\subsection{Corrections to the near-extremal limit} 
%

In our original effective microscopic theory we ignored brane curvature correction terms, assuming that,
\begin{eqnarray}
\left| \partial_\mu \vec{\Phi}_a \right|^2 \ll 1\,.
\end{eqnarray} 
We may now use our estimates to show that such terms are precisely negligible in the decoupling limit of \eqref{eq:decoupling}. We estimate,
 \begin{eqnarray}
\left| \partial_\mu \vec{\Phi}_a \right|^2 & \sim & \pi^2 T^2 \Phi^2 \nonumber \\
&\sim& \left( \frac{ N \mu \kappa^2 ( \pi T)^{n-2} }{ \Omega_{n-1} } \right)^{ \frac{2}{n-4} } \,,
\end{eqnarray} 
 and hence the approximation is consistent when, 
\begin{eqnarray}
\label{eq:coincident}
\left( \Omega_{n-1}^{-1}  \mu  \kappa^2 \pi^{n-2} T^{n-2} \right)^{\frac{2}{n-4}} \ll 1\,,
\end{eqnarray}
but this is precisely the same condition as we met earlier in equation \eqref{eq:omega} for the black hole to be in the decoupling limit.  
  
When $\left| \partial_\mu \vec{\Phi}_a \right|^2 \ll 1$ then $T \Phi \ll 1$ and hence we may say the separation of the branes in the transverse space, as measured by the scale $\Phi$, is small compared to the thermal scale. In this sense the branes may be thought of as being close together, or nearly coincident, relative to this thermal scale.

%
\section{Discussion and implications for holographic gauge theories\label{sec:discussion}}
%

We have considered black $p$-branes in supergravity. For total dimension $D = 10$ and $11$ this theory is appropriate to describe the various branes in string theory. However, we consider general $D$, where for $D<10$ the theory may be obtained by a consistent truncation of these cases after suitable dimensional reductions \cite{Duff:1994an}.

We have provided evidence for a microscopic description of these black $p$-branes in the near extremal limit which is a thermal bound state of elementary nearly coincident branes. We have argued that there is a classical $(p+1)$-dimensional world volume effective theory that is local and describes the dynamics of these branes. It is important that they are nearly parallel, and hence weakly curved, so that the theory is local in $(p+1)$-dimensions. We emphasise we have argued the form of this moduli theory, but it is something that in principle could be derived systematically in gravity using the classical effective field theory methods of \cite{Goldberger:2004jt}.

A key step is understanding that the horizon location $z_h$ is a physical quantity (rather than being coordinate dependent), giving the size of the horizon in the transverse space to the branes as measured in the brane metric. Then it naturally follows that we should identify 
the vev of the moduli fields with this horizon size $z_h$. 

We have estimated the finite temperature behaviour of the moduli theory by using the virial theorem. 
This naturally leads to the correct estimation of the energy density $\epsilon$ and the horizon size $z_h$ in terms of temperature. Perhaps the most surprising aspect of these estimates is that not only can we obtain the correct dependence of quantities on 
physical parameters, $N$, $\kappa$, $\mu$ and $T$, but we also may reproduce certain irrational factors, namely sphere volumes and factors of $\pi$.

The approximation that the branes are weakly curved, so that the gradients of moduli fields are small $\partial \Phi \ll 1$, is self consistent with our estimates  precisely when the thermodynamics reproduces that of $p$-branes in the decoupling or near-horizon limit. 
A simple physical picture is that the black $p$-brane is composed of nearly coincident elementary $p$-branes that are bound together at finite temperature in their mutual transverse space. When they are at sufficiently low temperature they behave as a warm strongly interacting liquid in this transverse space --- a warm $p$-soup. Heating this soup too much implies the branes are no longer nearly parallel, and become strongly curved, and do not describe the near extremal limit any more. Far from the extremal limit brane curvature corrections will not remain small, and we will not expect a local $(p+1)$-dimensional description of the system. 

If these ideas prove to be correct, then near extremal black $p$-branes may naturally be described in terms of a microscopic classical effective moduli field theory 
at finite temperature. This has an interesting implications, particularly for gauge/string duality which we now discuss.

One of the key aspects of the holographic duality between maximally supersymmetric $(p+1)$-dimensional Yang-Mills and IIA/B strings \cite{Maldacena:1997re,Itzhaki:1998dd}, 
as well as between ABJM gauge theory and M-theory \cite{Aharony:2008ug}, is that the classical moduli space of the gauge theory precisely describes parallel D$p$ or M$2$ branes. The dynamics of these moduli describe the low energy fluctuations of these branes.
Then quantum corrections to this classical moduli space, computed in an appropriate loop expansion, produce terms which include the gravitational interaction between branes \cite{Banks:1996vh,Becker:1997xw,Maldacena:1997re,Itzhaki:1998dd,Jevicki:1998ub} in our moduli theory.

Our moduli theory may be derived from the gauge theories by integrating out certain degrees of freedom. Following \cite{Wiseman:2013cda,Morita:2013wla} in Appendix \ref{app:SYM} we review precisely how this occurs at finite temperature for super Yang-Mills theory. More carefully our moduli theory is obtained as certain quantum corrections, when other terms are suppressed.
Integrating out degrees of freedom is exact. The moduli theory is obtained by truncating the suppressed terms (called the thermal and non-thermal corrections in the Appendix), and hence it should be regarded as a low energy effective theory derived from the gauge theory.
These loop corrections are computed for far separated branes, but precisely the ones that give our moduli action are believed to be protected by non-renormalisation theorems, and so also apply when the branes are close and interact strongly. 
As argued in \cite{Wiseman:2013cda}, the regime where we may consistently truncate to  our moduli theory is precisely the temperature range where the $\alpha'$ corrections to the  dual supergravity $p$-branes may be ignored. This is similar for the ABJM theory \cite{Morita:2013wla}.

It is highly non-trivial that our classical moduli theory of branes may be derived from gauge theory. However, as stated above, 
 the fact that the gauge theory reproduces weakly coupled brane dynamics and interactions, and that this is protected at strong coupling by non-renormalisation theorems,\footnote{Indeed the  non-renormalization theorem is so powerful that we can estimate the leading interaction terms even for the world volume theory of the M5-brane, where the  Lagrangian description is not known \cite{Morita:2013wla}.}
is one of the key historical pieces of evidence behind AdS/CFT and its generalisations. With this established, it follows rather naturally from the results in this paper that these gauge theories describe the thermodynamics of near extremal black $p$-branes when placed at finite temperature in an appropriate temperature regime.

Clearly it is important to determine if the $p$-soup picture is correct. If so, then it is important to understand if further calculations can tractably be performed using it. 
In order to do this, we must develop a better understanding of how quantities in the microscopic moduli theory are related to the semiclassical supergravity description. 
In this paper we have identified  the thermodynamic quantities between these descriptions, and proposed a map relating the horizon radius to the moduli vevs, but one in principle would like a more detailed dictionary. 

A direction that appears tractable is that for $p>0$ the branes have extended world-volume directions and an associated hydrodynamics \cite{Baier:2007ix,Bhattacharyya:2008jc}. Culinary experience suggests that $p$-soup may sometimes be only locally in thermal equilibrium, and it would be interesting to see if the hydrodynamics of black near extreme $p$-branes can be understood from our model.
Likewise, it would be interesting to understand rotating $p$-branes by `stirring' the soup. Also it would be interesting to explore whether one can extend the classical moduli theory to incorporate multiple varieties of branes and move beyond nearly parallel configurations.
%

%
\section*{Acknowledgements}
%

We would like to thank Andrew Hickling, Yoshifumi Hyakutake, Yoichi Kazama, Yoshinori Matsuo, Yuji Okawa, and David Tong for discussions. 
T.M. and S.S. would like to thank to the hospitality of the Theoretical Physics Group of Imperial College during their visit.
The work of T.M. was supported in part by Grant-in-Aid for Scientific Research (No. 24840046) from JSPS.
The work of S.S. is partially supported by Grant-in-Aid for JSPS fellows (No. 23-7749).

\appendix

%
\section{Appendix: supergravity $p$-branes in the decoupling limit}
\label{app:sugra}
%
Take total spacetime dimension $D$, and $p$-branes with $n$ transverse dimensions, so that $D = 1 + p + n$.
Then we start with the supergravity action;
\begin{eqnarray}
I = \frac{1}{2 \kappa^2} \int d^D x \sqrt{-g} \left( R - \frac{1}{2} (\partial \phi)^2 - \frac{1}{2 (p+2)!} e^{a\, \phi} F^2_{p+2} \right)
\end{eqnarray}
where the constant $a$ must take the particular value
\begin{eqnarray}
a^2 = 4 - \frac{2 (p+1)(n-2) }{D - 2}
\end{eqnarray}
in order to have asymptotically flat solutions. 
The asymptotically flat black $p$-brane is \cite{Duff:1994an}
\begin{eqnarray}
ds^2 &=&  \Delta_-^{\frac{n-2}{D-2}}  \left( - \frac{\Delta_+}{\Delta_-}  dt^2 +dx^i dx_i \right) + \Delta_-^{\frac{a^2}{2 (n-2)}} \left( \frac{1}{\Delta_+ \Delta_-} dr^2 + r^2  d\Omega^2_{n-1} \right)  \nonumber \\
e^{-2 \phi} &=& \Delta_-^a \nonumber \\
\Delta_{\pm} &=& 1 - \left( \frac{ r_\pm }{ r } \right)^{n-2}  \nonumber \\
F_{p+2} & = & r^{1-n} (n-2) \left( r_+ r_- \right)^{\frac{n-2}{2}}  \varepsilon_{p+2}
\end{eqnarray}
with $i= 1, \ldots , p$ and $\varepsilon_{p+2}$ is the volume form on $\mathbb{R}^{p+2}$ composed from the world volume directions and $r$.
The number of branes $N$ in this solution is given in terms of the charge density $Q$ as;
\begin{eqnarray}
N = \frac{Q}{\sqrt{2} \kappa \mu} = \frac{1}{2 \kappa^2 \mu} \int_{S^{n-1}} e^{a \, \phi} \star F = \frac{(n-2) \, \Omega_{n-1}}{2 \kappa^2 \mu}  \left( r_+ r_- \right)^{\frac{n-2}{2}} \; ,
\end{eqnarray}
and the mass density is,
\begin{eqnarray}
M = \frac{ \Omega_{n-1} }{2 \kappa^2}  \left( (n-1) r_+^{n-2} - r_-^{n-2} \right) \; ,
\end{eqnarray}
and entropy density;
\begin{eqnarray}
s & = &  \frac{2 \pi }{\kappa^2} \Omega_{n-1} r_+^{\frac{n-2}{2}} \left( r_+^{n-2} - r_-^{n-2} \right)^{\frac{n}{2 (n-2)}} \; .
\end{eqnarray}
In the extremal limit we have, $\sqrt{2} \kappa M = Q$.

To take the decoupling limit we must take the charge and mass to infinity, but keep the energy above extremity finite. Let us do this by defining,
\begin{eqnarray}
x = r_+^{\frac{n-2}{2}} \quad , \quad y = r_-^{\frac{n-2}{2}} \; ,
\end{eqnarray}
and then take $x \to \infty$, $y \to \infty$ such that $x/y = 1$ (this implies mass, charge $\to \infty$) but with $x - y \sim \frac{1}{x}$ to ensure the mass above extremity is finite.
Concretely we define new parameters;
\begin{eqnarray}
\label{eq:xy}
x = \frac{1}{\alpha} +  \frac{\lambda \alpha}{2} - \frac{\lambda^2 \alpha^3}{8} \; , \quad y = \frac{1}{\alpha} - \frac{\lambda \alpha}{2} + \frac{3 \lambda^2 \alpha^3}{8} \; ,
\end{eqnarray}
and take $\alpha \to 0$. Note that the parameterisation is picked so that; 
\begin{eqnarray}
M &=&  \frac{(n-2)\, \Omega_{n-1}}{2 \kappa^2} \frac{1}{\alpha^2} + \frac{n \Omega_{n-1}}{2 \kappa^2} \lambda - \frac{\lambda^2 \Omega_{n-1}}{2 \kappa^2}  \alpha^2 + O(\lambda^3\alpha^4) \nonumber \\
s & = &  \frac{2 \pi }{\kappa^2} \Omega_{n-1}  \left( 2 \lambda \right)^{\frac{n}{2 (n-2)}} \frac{1}{\alpha} + O(\alpha) \nonumber \\ 
Q &=& \frac{(n-2) \,\Omega_{n-1}}{\sqrt{2} \kappa} \frac{1}{\alpha^2} + O(\lambda^3\alpha^4)
\end{eqnarray}
so that mass and charge density indeed go to infinity, with the energy density above extremality, $\epsilon$, 
\begin{eqnarray}
\epsilon  \equiv M - \frac{1}{\sqrt{2} \kappa}Q =  \frac{n \Omega_{n-1}}{2 \kappa^2} \lambda  - \frac{\lambda^2 \Omega_{n-1}}{2 \kappa^2}  \alpha^2 + O(\lambda^3\alpha^4)
\end{eqnarray}
being finite, and importantly in this parameterisation the charge is held constant. Due to this we can deduce the temperature in the decoupled system as,
\begin{eqnarray}
T = \lim_{\alpha \to 0} \frac{d M}{d s} = \lim_{\alpha \to 0} \frac{d M/d \lambda}{d s /d\lambda} \; .
\end{eqnarray}
Now the number of branes is;
\begin{eqnarray}
N = \frac{(n-2)\, \Omega_{n-1}}{2 \kappa^2 \mu} \frac{1}{\alpha^2} + O( \alpha^2 ) \; .
\end{eqnarray}
To leading order in the $\alpha$ expansion we may relate the parameters $\lambda$ and $\alpha$ to the temperature and $N$ as,
\begin{eqnarray}
\label{eq:alpha}
\alpha^2 = \frac{n-2}{2} \frac{\Omega_{n-1}}{N \kappa^2 \mu} \; , \quad \lambda^{\frac{n-4}{n-2}} = \frac{2^{4+\frac{2}{n-2}} \pi^2 }{(n-2)^3 \Omega_{n-1}} N \kappa^2 \mu T^2 \; .
\end{eqnarray}
We now consider $n \neq 4$ and $n = 4$ separately.
\subsection{$n \neq 4$}
Using equation (\ref{eq:alpha}) we find, 
\begin{eqnarray}
\epsilon & = & 2^{\frac{3 n - 2}{n - 4}} n \left( \frac{ N \, \pi^2 \, T^2 }{ ( n - 2 )^3 }  \right)^{\frac{n-2}{n-4}}  \left( \frac{ \kappa^2  \mu^\frac{n-2}{2}   }{\Omega_{n-1}} \right)^{\frac{2}{n - 4}} \left( 1 -\frac{\lambda\alpha^2}{n}+ O\left( \lambda \alpha^2 \right)^2 \right)
\end{eqnarray}
where the leading deviations from extremality are shown and are controlled by small $\lambda \alpha^2$ given by, 
\begin{eqnarray}
\lambda \alpha^2 \sim \left(\Omega_{n-1}^{-1} N \kappa^2 \mu\, \pi^{n-2} T^{n-2}\right)^{\frac{2}{n-4}}.
\end{eqnarray}
where, as throughout this paper, `$\sim$' is correct up to and including factors of $\pi$ and $\Omega_{n-1}$.

Now we consider the decoupled geometry. If, in addition to \eqref{eq:xy}, we define a new radial coordinate $z$ as,
\begin{eqnarray}
\left( \frac{r_-}{r} \right)^{n-2} =  1 - \alpha^2 z^{n-2}
\end{eqnarray}
then to leading order in the decoupling limit $\alpha \to 0$ we obtain the (Einstein frame) metric;
\begin{eqnarray}
\label{eq:near-horizon}
ds^2 &=& \left(  \alpha^2 z^{n-2} \right)^{\frac{n-2}{D-2}} \left( - f dt^2 +  dx^i dx_i  \right) +  \left(  \alpha^2 z^{n-2} \right)^{- \frac{p+1}{D-2} } \left( \frac{1}{f} dz^2 + z^2 d\Omega^2_{n-1} \right) \nonumber \\
&=& \left( \frac{(n-2)\, \Omega_{n-1}}{2 \kappa^2 \mu N}   z^{n-2} \right)^{\frac{n-2}{D-2}} \left( - f dt^2 +  dx^i dx_i  \right) +  \left( \frac{(n-2)\, \Omega_{n-1}}{2 \kappa^2 \mu N}   z^{n-2} \right)^{- \frac{p+1}{D-2} } \left( \frac{1}{f} dz^2 + z^2 d\Omega^2_{n-1} \right) \nonumber \\
f & = & 1 - \frac{2 \lambda}{z^{n-2}}
\end{eqnarray}
with the dilaton profile,
\begin{eqnarray}
e^{-2 \phi} & = &  \left( \frac{(n-2)\, \Omega_{n-1}}{2 \kappa^2 \mu N}   z^{n-2} \right)^{a}.
\end{eqnarray}
Hence the position of the horizon $z_h$ is at $z_h^{n-2} = 2\lambda$ so yields the horizon position as;
\begin{eqnarray}
z_h = (2 \lambda)^{\frac{1}{n-2}} = \left( \frac{ 2^5  \pi^2 N T^2 \mu \kappa^2 }{(n-2)^3 \Omega_{n-1} } \right)^{\frac{1}{n-4}} \; .
\end{eqnarray}
In terms of this horizon position $z_h$ the energy density takes a simple form,
\begin{eqnarray}
\epsilon & = & \frac{n}{4} \frac{\Omega_{n-1}}{\kappa^2}  z_h^{n-2} \; .
\end{eqnarray}

\subsection{$n = 4$}
\label{app:hagedorn}
In the case $n=4$, $\lambda$ is not fixed as a function of temperature through equation (\ref{eq:alpha}).
It implies that the location of the horizon is a free parameter independent of temperature.
Instead, temperature is uniquely determined,
\begin{eqnarray}
\label{eq:TH}
T= \frac{1}{2\pi} \sqrt{ \frac{\Omega_3}{ N \mu \kappa^2}} \; .
\end{eqnarray}
Correspondingly the near horizon geometry (\ref{eq:near-horizon}) becomes the two-dimensional black hole \cite{Callan:1992rs, Maldacena:1997cg}.
It is then natural to parameterise the energy and entropy density as functions of the horizon location, $z_h$.

%
\section{Appendix: Scaling}
\label{app:scaling}
%

We will show that the effective theory (\ref{eq:eff-action}) 
admits a classical scaling.
This is presumably related to the generalised conformal symmetry in string theory \cite{Jevicki:1998ub}.
Assuming this classical scaling simply extends to the quantum theory, we can derive the supergravity temperature dependence of the scale $| \vec{\Phi} |$ and the entropy density $s$.

Under a scaling,
\begin{eqnarray}
\tau & \to & \Lambda^{-1} \tau  \; , \quad
x^i  \to  \Lambda^{-1} x^i \; , \quad
\vec{\Phi}_a \to \Lambda^{\frac{2}{n-4}} \vec{\Phi}_a 
\end{eqnarray}
then the classical Euclidean action \eqref{eq:eff-action} together with the on-shell terms \eqref{eq:onshell} scales as,
\begin{eqnarray}
 S^{E} \to \Lambda^{1 - p + \frac{4}{n-4}}  S^{E} \; .
\end{eqnarray}
For such a scaling regime to occur it is clear that all the on-shell higher derivative terms in the action will be active, and hence if this classical scaling extends to the quantum theory, one would expect it to apply in the strongly coupled regime.
Let us assume that there is a temperature range in the theory where this scaling is manifest. Then this implies that under a scaling of temperature $T$ the thermal expectation value of the scalars, and the entropy density $s$ (which we assume scales as the  Euclidean action density) scale as,\begin{eqnarray}
\label{eq:scaling}
T \to \Lambda T \; \implies \quad \left\langle | \vec{\Phi} | \right\rangle \to \Lambda^{\frac{2}{n-4}} \left\langle | \vec{\Phi} | \right\rangle \; , \quad s \to \Lambda^{2 + \frac{4}{n-4}} s \; ,
\end{eqnarray}
and hence in such a scaling regime the temperature dependence of these quantities is determined as,
\begin{eqnarray}
 \left\langle | \vec{\Phi} | \right\rangle \propto T^{\frac{2}{n-4}}  \; , \quad s \propto T^{2 + \frac{4}{n-4}} \; .
\end{eqnarray}
We immediately note that this scaling reproduces the thermal dependences seen in the supergravity quantities, after the identification that $z_h \sim \langle |\vec{\Phi}_a| \rangle$ as seen in equations \eqref{eq:sugraentropy} and \eqref{eq:zh}. 

The appearance of this scaling property is natural, since the virial theorem tells us that $S \sim S_2 \sim S_4 \sim S_{on-shell, 2l} $.
Thus these terms possess the same temperature dependence and exhibit the scaling (\ref{eq:scaling}).
We emphasise this scaling depends on strong coupling. 
At weak coupling then $S_2$ will dominate and we expect,
\begin{eqnarray}
\label{eq:free}
 \left\langle | \vec{\Phi} | \right\rangle \propto T^{\frac{p-1}{2}}  \; , \quad s \propto T^{p}  \; ,
\end{eqnarray} 
and hence we have a very different scaling with temperature.

%
\section{Appendix: Maximally supersymmetric Yang-Mills}
\label{app:SYM}
%

In this appendix we discuss the D$p$ brane case where a fundamental description of the decoupling limit is given in terms of maximally supersymmetric Yang-Mills (SYM). We confirm that our microscopic theory is reproduced from SYM in a certain limit, and we consider the corrections to this effective action, and when they are negligible. We note that the fact that the SYM reproduces the weak coupling gravitational interaction between D$p$ branes is key evidence historically behind the conjectured duality \cite{Itzhaki:1998dd,Jevicki:1998ub}.

SYM in $p$ spatial dimensions has the action,
\begin{eqnarray}
{S}_{YM} = \frac{1}{g_{YM}^2}  \int dt dx^p \, \tr\left[  - \frac{1}{4} F_{\mu\nu}^2 - \frac{1}{2} D^\mu \underline{\underline{\phi}}^I D_\mu \underline{\underline{\phi}}^I + \frac{1}{4} \left[ \underline{\underline{\phi}}^I , \underline{\underline{\phi}}^J \right]^2  \right] + \mathrm{fermions}
\end{eqnarray}
and we take the gauge group $U(N)$. The scalar fields $\underline{\underline{\phi}}^I$ and gauge field $\underline{\underline{A}}^\mu$ are $N \times N$ hermitian matrices transforming in the adjoint of the gauge group, with indices $I = 1, 2,  \ldots , 9-p$ and $\mu = 0,1,\ldots, p$. For general $p$ the gauge coupling has dimensions,
\begin{eqnarray}
\left[ g_{YM}^2 \right] = 3 - p \quad \implies \quad \left[ \underline{\underline{\phi}} \right] = 1  \; .
\end{eqnarray}
The classical vacua are gauge equivalent to configurations where $\underline{\underline{A}}^\mu$ and $\underline{\underline{\phi}}^I$ are both constant and diagonal. Given this classical vacuum moduli space, we may prompte the constant diagonal values to slowly varying  degrees of freedom - the moduli -  $A^I_a(t, x)$ and $\phi^I_a(t, x)$. Then we have,
\begin{eqnarray}
( \underline{\underline{A}}^\mu )_{ab} = A^\mu_a(t, x) \delta_{ab} \; , \quad  ( \underline{\underline{\phi}}^I )_{ab} = \phi^I_a(t, x) \delta_{ab} \; .
\end{eqnarray}
Thinking of $\phi^I_a$ as a vector in the $\mathbb{R}^n$ transverse space, we use the notation $\phi^I_a = \vec{\phi}_a$.

Now consider the theory at finite temperature $T$ using the Euclidean formulation. As shown in \cite{Wiseman:2013cda} the classical action for the moduli takes the form,
\begin{eqnarray}
\label{eq:moduli}
{S}^{E,classical} =  \frac{1}{g_{YM}^2} \int d\tau dx^p  \sum_{a}  \left( \frac{1}{2} \partial^\mu \vec{\phi}_a\cdot\partial_\mu \vec{\phi}_a + \frac{1}{4} F_{\mu\nu a} F^{\mu\nu}_a \right) \; ,
\end{eqnarray}
where $F_{\mu\nu a} = \partial_\mu A_{\nu a} - \partial_\nu A_{\mu a}$ are the field strengths for the $N$ $U(1)$ gauge fields. 
Since the scalar moduli are uncharged, we will largely ignore the gauge field moduli for the following.
\\

\noindent
{\bf The moduli theory of branes}

The action for the scalars above looks similar to the first term in our moduli theory of nearly parallel branes. We now discuss how the interaction term arises from the full Yang-Mills theory.

Integrating out the off diagonal degrees of freedom of the gauge and scalar matrix fields, one generates various corrections to this classical moduli action. 
The loop corrections may be characterised as those with no explicit temperature dependence, which we term non-thermal corrections, and those with explicit temperature dependence, which we term thermal. 
The terms that give rise to our effective microscopic theory we term the \emph{leading} terms. We define the leading terms as the \emph{lowest} derivative \emph{non-thermal} terms (i.e. those generated at zero temperature).

Thus the leading terms that generate our moduli theory derive from the non-thermal terms. Due to the supersymmetry there is no potential or correction to the two derivative classical action generated. The leading terms at 1-loop have four derivative and are,
\begin{eqnarray}
{S}^{E,1-loop}_{leading} &=&  - \int d\tau dx^p \sum_{a<b}  \frac{\Gamma\left( \frac{7 - p}{2} \right)}{( 4 \pi )^\frac{1+p}{2}} \Bigg( 
2 \frac{ \left( \partial_\mu \vec{\phi}_{ab} \cdot \partial_\nu \vec{\phi}_{ab} \right) \left( \partial^\mu  \vec{\phi}_{ab}  \cdot \partial^\nu  \vec{\phi}_{ab}  \right) }{ |  \vec{\phi}_{ab}  |^{7-p} } 
\nonumber \\
&& \qquad \qquad \qquad \qquad
\qquad \qquad \qquad \qquad
-  \frac{\left( \partial_\mu  \vec{\phi}_{ab}  \cdot \partial^\mu  \vec{\phi}_{ab}  \right)^2 }{ |   \vec{\phi}_{ab}  |^{7-p} }
\Bigg)  + \ldots
\end{eqnarray}
where we use the notation $ \vec{\phi}_{ab} \equiv \vec{\phi}_a - \vec{\phi}_b$. For D$p$-branes the brane tension $\mu$ and gravitational coupling $\kappa^2$ are given in terms of the string coupling $g_s$ and $\alpha'$ as;
\begin{eqnarray}
\mu = (2 \pi)^{- p} g_s^{-1} \alpha'^{-\frac{1+p}{2}} \; , \quad 2 \kappa^2 = (2 \pi)^7 g_s^2 \alpha'^4 \; .
\end{eqnarray}
In the decoupling limit the Yang-Mills coupling $g_{YM}^2$ is given in terms of string theory quantities as,
\begin{eqnarray}
g_{YM}^2 = \frac{1}{( 2 \pi \alpha')^2 \mu} \; .
\end{eqnarray}
Identifying our general moduli $\vec{\Phi}_a$ with these scalar moduli $\vec{\phi}_a$ the two derivative terms imply the relation,
\begin{eqnarray}
\sqrt{\mu} \vec{\Phi_a} = \frac{1}{g_{YM}} \vec{\phi}_a \; .
\end{eqnarray}
Then written in the $\vec{\Phi}_a$ normalisation the 4 derivative 1-loop non-thermal term becomes,
\begin{eqnarray}
{S}^{E,1-loop}_{leading} 
&=&  - \left(\frac{1}{4 (7-p)}\right) \frac{\mu^2 \kappa^2}{Vol\left( S_{8-p} \right)}\\
&& \times \int d\tau dx^p \sum_{a<b}  \Bigg( 
\frac{ 2 \left( \partial_\mu \vec{\Phi}_{ab} \cdot \partial_\nu \vec{\Phi}_{ab} \right) \left( \partial^\mu  \vec{\Phi}_{ab}  \cdot \partial^\nu  \vec{\Phi}_{ab}  \right) 
- \left( \partial_\mu  \vec{\Phi}_{ab}  \cdot \partial^\mu  \vec{\Phi}_{ab}  \right)^2 }{ |   \vec{\Phi}_{ab}  |^{7-p} } \Bigg)  + \ldots \nonumber
\end{eqnarray}
and hence agrees precisely with \eqref{eq:eff-action} with the correct assignment of transverse dimensions, $n = 9 - p$.

Beyond 1-loop rather little is known about the structure of this moduli theory. Certain results are known for 2-loop contributions for $p=0$ \cite{Okawa:1998pz} and $p=3$ \cite{Buchbinder:2001ui}. We note that these are consistent with our expectation for the on-shell higher gravitational interactions in \eqref{eq:onshell}. Again we emphasise that this consistency was a strong historical motivation for believing the holographic correspondence, and also Matrix theory (see for example \cite{Becker:1997xw,Okawa:1998pz}).

\subsection{Corrections to the effective microscopic theory}

Thus our microscopic moduli action is reproduced by the leading terms at 1-loop. However there are other terms that have been ignored. We now describe them, and check within our estimating scheme when it is consistent to ignore them.

The corrections to the leading term at 1-loop are of two varieties.\footnote{
In addition to these two corrections, different
types of corrections, i.e. monopole corrections at $p=2$, may appear
if we do not take the 't Hooft limit.
}
Firstly the higher derivative non-thermal terms, and secondly the thermal terms that explicitly depend on temperature.
\begin{itemize}
\item
{\bf Non-thermal higher derivative corrections}

We believe the leading non-thermal terms are corrected at higher derivative order schematically as,
\begin{eqnarray}
 S^{E,1-loop, q}_{non-thermal}   & \sim &   \int d\tau dx^p  \sum_{a < b}^{}  \left( \frac{ \left( \partial \vec{\phi} \right)^{4} }{ \left| \vec{\phi} \right|^{ 7 - p } } \right) \left( \frac{ \left( \partial \vec{\phi} \right)^{2} }{ \left| \vec{\phi} \right|^{4} } \right)^q \; ,
  \end{eqnarray}
for integer $q > 0$, where we ignore non-dimensional factors. 

It is a very important point that the non-thermal terms at 1-loop generate a leading term with $4$ derivatives and higher derivative corrections. Therefore they do not effect the 0-loop leading term. This is due to the form of the 2 derivative term being protected by supersymmetry. It is expected that at higher loops the 4 derivative leading term is also protected from non-thermal corrections by non-renormalisation theorems (see for example \cite{Paban:1998ea,Paban:1998qy,Lowe:1998ch,Sethi:1999qv,Taylor:2001vb} for $p=0$).

\item
{\bf Thermal corrections}

In addition to the non-thermal terms at 1-loop that only depend on temperature due to the integral over Euclidean time, there are also 1-loop correction terms that have explicit temperature dependence. These were computed in \cite{Wiseman:2013cda}. Unlike the non-thermal terms, these are not required to have world volume Lorentz invariance. They correct all derivative orders, but importantly appear to be exponentially suppressed in $\beta | \phi_{ab} |$. Hence under appropriate conditions all these terms are irrelevant in the scaling regime. 

At one loop the thermal correction gives contributions to all even derivatives. The zero derivative term generates a potential (in the case $p=0$ that computed in \cite{Ambjorn:1998zt}), which to leading order when $\beta | \phi_{ab} | \gg 1$ (so that the terms are exponentially suppressed) is of the form,
\begin{eqnarray}
\label{eq:thermal}
{S}^{E,0}_{thermal} & = &  - \frac{16}{ (2 \pi)^{p/2} } \int d\tau dx^p \sum_{a<b}   \frac{ U_a U^\star_b + U_b U^\star_a }{ \beta^{1 + p} }  e^{- \beta |  \vec{\phi}_{ab} |} \left( \beta |  \vec{\phi}_{ab} | \right)^{p/2}  \; ,
\end{eqnarray}
where $U_a$ is the Polyakov loop around the Euclidean time circle, so that $U_a = e^{i \oint d A_a} = e^{i \oint d \tau A^0_a}$. At zero derivatives this term is world volume Lorentz invariant. At higher derivative orders analogous exponentially suppressed terms in $\beta | \phi_{ab} | \gg 1$ arise, correcting the two, four and higher derivative terms in the moduli scaling action.

Beyond one loop we do not know of a calculation of these thermal corrections, but we expect that they are also thermally suppressed by positive powers of $e^{- \beta |  \vec{\phi}_{ab} |}$.

\end{itemize}

Hence there are many terms that correct the moduli action composed of the leading terms at 1-loop, but these are irrelevant if the two conditions,
\begin{eqnarray}
\frac{ \left( \partial \vec{\phi} \right)^{2} }{ \left| \vec{\phi} \right|^{4} } \ll 1 \; , \qquad e^{- \beta |  \vec{\phi}_{ab} |} \ll 1 
\end{eqnarray}
hold. We now show that using our estimates, it is true that both these conditions can be met and give a condition on the temperature which precisely coincides with the condition that the supergravity thermodynamics is valid, and not corrected by stringy effects.

\subsection{Estimates}

The condition \eqref{eq:coincident} is automatically satisfied in the decoupling limit of D$p$-branes. However in a full string embedding of the supergravity one must still make sure that the supergravity is a good approximation.
In \cite{Itzhaki:1998dd} it was shown that this is true at large $N$ provided we can ignore $\alpha'$ corrections which implies that,
\begin{eqnarray}
 \frac{T}{ \left( N g_{YM}^2 \right)^{\frac{1}{3-p}}}   \ll 1 \label{eq:alpha2} \; .
\end{eqnarray}
Using our estimates, and translating to the variable $\vec{\phi}_a$ we have,
\begin{eqnarray}
\Phi &\sim& \left( \frac{ N \mu \kappa^2 \pi^2 T^2}{ \Omega_{n-1} } \right)^{ \frac{1}{n-4} } \quad \implies \qquad \phi \sim \left( \frac{ N g_{YM}^2  \pi^{6-p} T^2}{ \Omega_{8-p} } \right)^{ \frac{1}{5-p} } \; .
\end{eqnarray}
For the non-thermal and thermal corrections to be negligible we require,
\begin{eqnarray}
\frac{ \left( \partial \vec{\phi} \right)^{2} }{ \left| \vec{\phi} \right|^{4} } \ll 1 \; , \quad \beta \phi \gg 1 
\end{eqnarray}
respectively. However we note that our estimate for the first,
\begin{eqnarray}
\frac{ \left( \partial \vec{\phi} \right)^{2} }{ \left| \vec{\phi} \right|^{4} } \sim \frac{\left( \pi T \phi \right)^2}{\phi^4} \sim \frac{1}{\left( \beta \phi \right)^2 } \ll 1
\end{eqnarray}
just reproduces the second condition. Thus both corrections are small provided $\beta \phi \gg 1$. Using our estimate above we see,
\begin{eqnarray}
\beta \phi \gg 1 \quad \implies \qquad \frac{T}{ \left( N g_{YM}^2 \right)^{\frac{1}{3-p}}}   \ll 1 \; ,
\end{eqnarray}
and hence we see that the corrections to the effective microscopic action become important precisely when string $\alpha'$ corrections become important for the black hole. As noted in \cite{Wiseman:2013cda} while our estimates predict the thermodynamic behaviour of black holes, the fundamental theory also provides a mechanism to break this by corrections to the effective moduli theory, which correspond to the break down of the supergravity solution.

%
\bibliographystyle{JHEP}
\bibliography{paperV1}

%

\end{document}